\documentclass[a4paper,fleqn]{cas-dc}

\usepackage[numbers]{natbib}
\usepackage{graphicx}      % For including images with \includegraphics
\usepackage{amsmath,amssymb,amsfonts} % Essential for math environments, symbols, and fonts
\usepackage{float}

\usepackage{siunitx}       % Proper formatting of units and numbers
\usepackage{fancyhdr}      % Custom headers and footers

\usepackage{lipsum}        % Generates placeholder text (for testing layouts)

\usepackage{tabularx}	   % advanced tables
\usepackage{booktabs}	   % Additional table lines
\usepackage{xcolor}        % For using colors in text, tables, or figures
\usepackage{multirow}	   % Allows multi row cells in tables

\usepackage{hyphenat}
\def\tsc#1{\csdef{#1}{\textsc{\lowercase{#1}}\xspace}}
\tsc{WGM}
\tsc{QE}
\begin{document}
	\let\WriteBookmarks\relax
	\def\floatpagepagefraction{1}
	\def\textpagefraction{.001}
	%\let\shorttitle\relax 
	
	\shortauthors{Brodie et al.}   
	
	\title [mode = title]{Design and Benchmarking of the Data Distribution Service for Real-Time Interoperable Agricultural Machinery Communications}
	\shorttitle{}
	
	%       gplus=<gplus id>] 85354 Freising
	
	\author[1]{Samuel Brodie}

	\affiliation[1]{organization={Technical University of Munich, Chair of Agrimechatronics and Munich Institute of Robotics and Machine Intelligence (MIRMI)}, 
		city={85354 Freising},
		state={Bavaria},
		country={Germany}}

	\author[1]{Henri Hornburg}
	\author[1]{Daniel Ostermeier}
	\author[1]{Maksim Pavlov}
	\author[1]{Timo Oksanen}
	\cormark[1]
	\ead{timo.oksanen@tum.de}

%	\linenumbers
	
%	\switchlinenumbers
	
	\cortext[1]{Corresponding author}
	
	
	
   % Abstract 
   \begin{abstract}
Inter-manufacturer plug-and-play communication in agricultural machinery is currently based on the ISO 11783 standard series, which specifies a 250 kbit/s CAN bus communication layer. To support higher-bandwidth use cases, the ISO~23870 series is being developed for next-generation Ethernet-based agricultural machine-to-machine communication. However, modern Ethernet/IP-based architectures often make use of a middleware for discovery, data exchange, quality of service configuration, and security. This paper evaluates the Data Distribution Service (DDS) as a candidate middleware for secure, plug-and-play agricultural machinery networking. A DDS-based proof-of-concept communication design is presented for a representative Task Controller (TC) and implement scenario, including implement-description topics and separate best-effort and reliable topics for runtime process data. The design was implemented in C++ using the FastDDS library and benchmarked on embedded hardware representative of agricultural machinery. Runtime throughput was evaluated for one-to-one and one-to-two TC--implement scenarios under four DDS security configurations. The results show that DDS security mechanisms substantially reduce maximum throughput on embedded hardware. In the tested best-effort scenarios, signing and encryption reduced mean throughput by approximately 70---84\% compared with the unsecured configuration. Nevertheless, the encrypted one-to-one best-effort case achieved approximately 4980 received process data updates per second on both the TC and implement, corresponding to about 50 process data updates per second per simulated section for 100 rate-controllable sections. These results indicate that DDS is a technically plausible middleware candidate for secure Ethernet-based agricultural machinery interoperability, while further work is required to evaluate latency, scalability, vendor interoperability, and lower-power devices.
   \end{abstract}

%\nocite{*}

% Keywords
\maketitle

%===============================================================================

\section{Introduction}
The ISO 11783 standard series defines a manufacturer-independent, plug-and-play communications interface for agricultural machines; it enables various functionalities such as automatic implement control and user interface uploads. ISO 11783 has been standardised since 2001 \cite{02_buzzword} and uses 250 kbit/s CAN bus \cite{01_ISO}. A new standard series, ISO 23870, is under development to address the industry's future plug-and-play use cases, such as improved Task Controller (TC), camera integration \cite{03_dcs}, advanced automation, and modern cybersecurity. 

ISO 23870 is being developed as a next-generation successor to ISO 11783 and standardises Ethernet and IP for the lower OSI layers. The added complexity of such a network means that instead of standardising individual Ethernet/IP packets (as is done with CAN in ISO 11783), the complexity can be better managed by using a standardised middleware layer to benefit from the large community of products and vendors which exist to solve similar connectivity challenges. Middleware is a class of generic software components which typically supports standardised interfaces and protocols, and enables distributed computing \cite{04_middlewareDef}.

Various middlewares exist, and each has unique benefits and drawbacks. One such middleware to be considered is the Data Distribution Service (DDS), a middleware standardised by the Object Management Group (OMG) for secure, interoperable, data-centric communication \cite{07_ddsSpec, 08_rtpsSpec, 09_ddsSecuritySpec}. However, existing DDS benchmarks commonly forego enabling the required security configuration, and the architectures tested are not specific to agricultural use cases.

The TC is a software device which uses internal proprietary algorithms to send setpoints to the implement to adjust the application rate or turn sections on/off and the TC also logs data from the implement \cite{06_iso10}. Therefore, the communications network in next-generation agricultural machines must be secured. The data generated by TCs and implements can include yield and fertilisation strategies, data that is secretive in nature. Furthermore, next-generation autonomy messaging, while not necessarily secret, must be secured against message spoofing and tampering which would result in unsafe behaviours. 

\begin{figure*}
   \centering
   \includegraphics[width=0.75\linewidth]{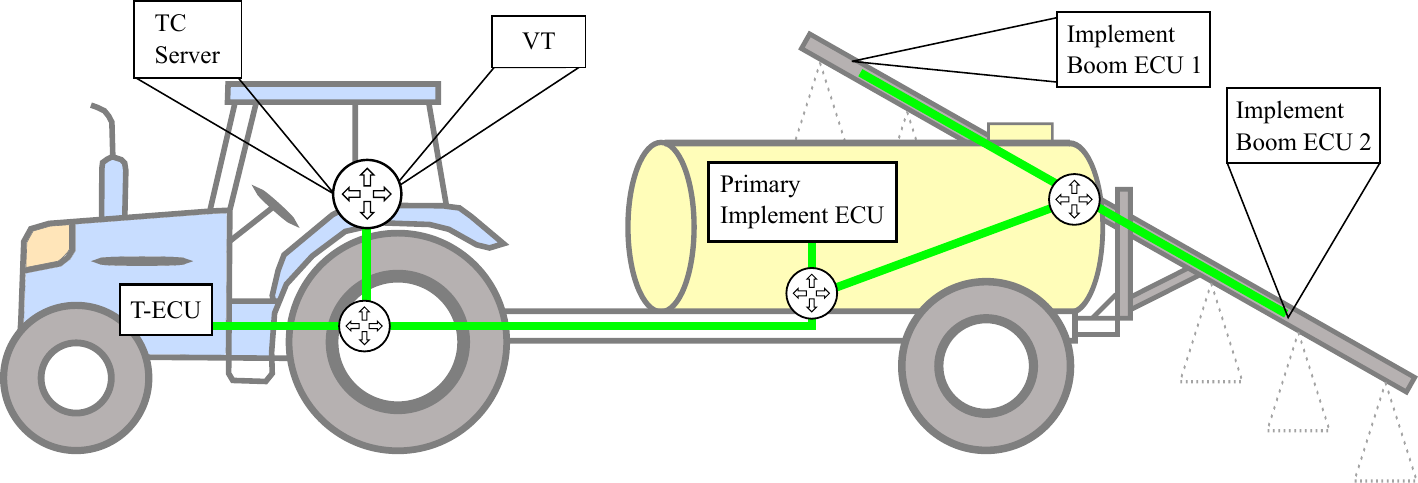}
   \caption{A typical use case for next-generation agricultural machinery networking. An Ethernet network connecting a tractor (T-ECU), implement, Virtual Terminal (VT), and Task Controller (TC), which could all be from different manufacturers. The topology uses peer-to-peer connections via Ethernet switches.}
   \label{fig:typical_network}
\end{figure*}

This work investigates the suitability of DDS for development of a secure, next-generation, wired, machine-to-machine interoperability network for agricultural machinery (such as that shown in Figure~\ref{fig:typical_network}, for example). To this end, a system design is presented that enables plug-and-play interoperable sharing of process data between implements and a TC without prior configuration. A proof-of-concept (PoC) implementation of the system is then benchmarked using embedded hardware typical of agricultural machines and the impact of different security policies is shown.

The research questions addressed by this work are:
\begin{itemize}
    \item Can DDS be used to support \textit{ad hoc}, cross-vendor TC–implement communication in an agricultural networking scenario?
    \item What is the runtime throughput (in terms of process data updates per second) of the proposed DDS design when running on embedded hardware?
    \item How do different DDS security configurations affect this throughput?
\end{itemize}

\section{DDS background}
DDS nodes (\textit{DomainParticipants}) discover one another dynamically on the network in a peer-to-peer manner with the standardised Simple Discovery Protocol (SDP) without requiring prior knowledge of the network or a central broker. Communication is organised around topics, a named data channel associated with a specific data type and, optionally, a topic partition. \textit{DomainParticipants} exchange data samples through endpoints (\textit{DataWriters} and \textit{DataReaders}), each of which is bound to a single topic. 

From the application perspective, communications are decoupled: Data samples associated with a particular topic are delivered to the relevant readers and applications interact with topics solely by sending data to or reading data from their local endpoint, without requiring direct awareness of the remote endpoints on other devices. Quality of Service (QoS) profiles can affect network-level messaging by setting data transfer to reliable, or best-effort, or configuring the data update frequency, again, decoupled from the application logic.

The OMG DDS Security Specification defines a security model and a Service Plugin Interface as well as built-in interoperable plugins for authentication, access control, and cryptography \cite{09_ddsSecuritySpec}. The security configuration is defined in each \textit{DomainParticipant}'s Governance Document. The security is robust \cite{10_ddsSecAnalysis}, however, metadata such as topic names and permissions can be leaked to eavesdroppers \cite{11_ddsSecIssue}.

\textit{DomainParticipants} also have a Permissions Document which is signed by a central root Permissions CA and linked to the \textit{DomainParticipant}'s identity. The permissions document is shared during discovery to inform remote \textit{DomainParticipants} of the rules and actions that the local \textit{DomainParticipant} should be allowed to do in the domain on a per-topic basis.

\section{Related work}
\subsection{Other DDS-based standards}
DDS is well established and is the middleware for several prominent projects:
\begin{itemize}
    \item Robot Operating System 2 (ROS 2) \cite{14_ros2Pub}.
    \item NATO Generic Vehicle Architecture \cite{15_NGVA5}.
    \item A DDS binding is available for AUTOSAR Adaptive Platform \cite{16_autosar1, 17_autosarDdsWhitepaper}.
\end{itemize}

DDS acts as both the transport and security layer in Robot Operating System 2 (ROS 2) \cite{14_ros2Pub}. However, interoperability between different DDS vendors is not a priority in ROS 2 and it is common for a single DDS library to be used for all parts of the system. Additionally, the ROS 2 community identified several drawbacks such as: poor scaling of discovery network traffic, issues with large data transfers, and degraded performance over wireless networks \cite{18_ros2Alternatives}. As such, there is a proposal to use Zenoh as an alternative ROS 2 middleware instead of DDS \cite{18_ros2Alternatives}.

The NATO Generic Vehicle Architecture uses DDS for interoperable communications between vehicle components \cite{15_NGVA5} and the standard defines the DDS topic names and data types that are used as interfaces. In addition, the QoS policies are standardised. The Generic Vehicle Architecture security standard is not publicly available. The NATO Generic Vehicle Architecture is not for "hot" plug and play, but rather for improved system integration and modularity of components.

The AUTOSAR Adaptive Platform defines an API and expected behaviours for application developers \cite{16_autosar1}. It is flexible and allows for different communications protocols to be used as its middleware, of which DDS is one \cite{17_autosarDdsWhitepaper}. The Specification of DDS Service Discovery Protocol document defines methods for implementing service discovery for AUTOSAR Adaptive Platform via DDS \cite{19_autosar_dds}. 

\subsection{Utilisation of DDS}
Wagner, et al. \cite{20_ddsCustomPlugins} are the first to implement custom security plugins by extending the built-in Authentication and Access Control plugins to utilise Trusted Platform Module-based remote attestation. 

Wu, et al. \cite{21_drivingMiddleware} propose a benchmark to evaluate a middleware’s performance with regards to usage in autonomous vehicles (e.g., for sensor interconnection) and show the performance impacts of using DDS middleware with UDP rather than shared memory communication. The UDP-based, data-centric design of DDS benefits low-power, wirelessly connected IoT devices \cite{22_sdn}, however, there are differences in performance of different vendors' DDS implementations, with \cite{23_bode_dds} finding RTI Connext to be the best overall with FastDDS and CycloneDDS also performing well. Enabling the security features of ROS 2 has also been shown to negatively impact runtime \cite{13_ros2Perf} performance.

The open-source project OpenICE uses DDS as its communication middleware to implement the Integrated Clinical Environment standard, AAMI 2700-1 for interoperable communications for medical equipment \cite{24_openICE, 25_openIce1}.

Time-sensitive networking (TSN) is a group of IEEE 802.1 standards which define methods of deterministic transport \cite{26_tsnIntro}. TSN requires time synchronisation between all network devices \cite{26_tsnIntro}. Lu, et al. \cite{27_ddsInTsn} present TS-DDS which combines DDS and TSN for use with AUTOSAR Adaptive Platform and maps DDS QoS policies onto TSN policies, while Zhang, et al. \cite{28_ddsTsn} present a system for the AUTOSAR Classic Platform which combines TSN and a lightweight DDS for automotive using a combined QoS scheme. In both works the TSN network is pre-configured, not plug-and-play \cite{27_ddsInTsn, 28_ddsTsn}.

Several works have addressed the discovery process of DDS to improve its scalability, in particular by employing Bloom filters \cite{29_bloomDiscovery,30_ddsNodeDiscovery,31_madDDS} or by utilising a so-called "discovery server" to broker the data such as the eProsima Discovery Server \cite{32_fastDdsDiscovery} and Real-Time Innovations Inc. Discovery Service \cite{33_rtiDiscoveryServer}. Both Bloom filters and discovery servers can reduce the amount of discovery data shared by filtering \textit{DomainParticipant} announcement messages so that \textit{DomainParticipants} only proceed through the complete discovery process with one another if they need to further communicate post-discovery. For example, if two \textit{DomainParticipants} have no topics in common then they will not need to proceed to the Endpoint Discovery Phase. Enabling security during discovery has also been shown to reduce performance \cite{12_rtsSecBenchmarks}.

\subsection{Research gaps}
Middlewares such as OPC UA have been researched for agricultural machinery and ISO 11783 Task Controller \cite{05_mattiOpcua}. However, no reported studies were found that address the use of DDS in plug-and-play systems of this nature where a non-technical end user connects two or more devices in the field. No benchmarks have been published for this specific agricultural interoperability scenario.

DDS can be configured to implement security at Layer 7, which simplifies implementation but requires computation of cryptographic material for each message.

\section{DDS system design for next-generation networking}
% Design and development
\label{sec:systemDesign}
\subsection{Plug-and-play communications}
For plug-and-play interoperability between machines, the topic names and data types are predefined because the physical meaning of each value must be standardised and known at design time. This is in contrast to a machine-to-human interface where the topic names and data types can themselves impart information.

The design assumes that devices connecting to the network are able to obtain an IP address before the middleware discovery begins. The Identity CA and Permissions CA roles should be fulfilled by a trusted industry body to sign and distribute the security artefacts. The ISO NAME concept from ISO 11783 is maintained, whereby each machine has a unique 64-bit identifier.

\subsubsection{DDOP equivalent}
\label{sec:ddopEquiv}
DDS does not allow modelling of parent-child relationships between data objects natively. However, this type of relationship is required for maintaining at least the same modelling functionality as in the ISO 11783 DDOP, which is a hierarchy of parent and child elements with signals (called DDIs in ISO 11783) attached to them. Therefore, this parent-child relationship needs to be defined on the application layer. 

To transfer the equivalent of the ISO 11783 DDOP, two DDS topics are defined, the elements and their parental relationships are written to one topic, and the list of signals supported by each element is written to the \textit{t\_signals\_linking} topic. An example of the signal linking is shown in Table~\ref{tab:ddiLinks}. 

The DDOP can be automatically uploaded to the server by setting the DURABILITY QoS to TRANSIENT\_LOCAL in both topics without the need for a periodic TC-status message as in ISO 11783.

\begin{table*}[width=.9\textwidth,cols=3,pos=h]
	\caption{Example linking of signals to DDOP element numbers in the \textit{t\_signals\_linking} topic. The implement is able to report its working height and fill level in the root element, and has 3 sub units with independently controllable rates.}
	\label{tab:ddiLinks}
	\begin{tabularx}{.6\textwidth}{|X|p{0.15\textwidth}|X|}
		\hline
		\textbf{Element} & \textbf{Handling Group} & \textbf{Unit} \\
		\textbf{Reference} & ~ & ~ \\
		\hline
		NAME: 0xFF0001 & Working height & Numerator: $m$ \\
		Element ID: 100 & ~ & Denominator: None\\
		\hline
		NAME: 0xFF0001 & Tank fill level & Numerator: $\%$ \\
		Element ID: 100 & ~ & Denominator: None\\
		\hline
		NAME: 0xFF0001 & Application rate & Numerator: $kg$ \\
		Element ID: 201 & ~ & Denominator: $m^2$ \\
		\hline
		NAME: 0xFF0001 & Application rate & Numerator: $kg$ \\
		Element ID: 202 & ~ & Denominator: $m^2$ \\
		\hline
		NAME: 0xFF0001 & Application rate & Numerator: $kg$ \\
		Element ID: 203 & ~ & Denominator: $m^2$ \\
		\hline
	\end{tabularx}
%	\vspace{2mm} % add space to bottom of last table of group
\end{table*}

\subsubsection{Process data}
After the DDOP is uploaded, the TC server has the necessary information to begin control, i.e., the implement can notify that it is ready to work, and the TC server can begin to send setpoints, and receive actuals.

Some of the process data values shared between the TC server and implement such as application rates are generally sent with a high update frequency. Other process data values, such as geometry values, are sent infrequently and it is important to ensure that they are received correctly. This split between frequently updated and rarely updated leads to a split of needing some values to be sent with best-effort and some values to be sent reliably. Since best-effort and reliable are incompatible QoS values, the design separates process data into two topics with different QoS settings. These topics are called \textit{t\_pd\_values\_best\_effort} and \textit{t\_pd\_values\_reliable}. This is done to avoid the system being designed to contain endpoints with mismatched QoS which fail to connect and make future troubleshooting more complex.

An example of the transfer of the process data variables that may be shared on the best-effort process data topic is shown in Table~\ref{tab:t_pd_values_example}; the data types are the same for reliable process data. The element reference of the controlled/reported process acts as the key field, the handling group enumeration indicates the real-world process attached to the value, the units field represents what type of units are used (e.g., mass or volume) and the value itself, whose interpretation depends on the handling feature (e.g., setpoint, actual, minimum). The size of a process data update is 40 Bytes at the application layer.

\begin{table*}[width=.9\textwidth,cols=5,pos=h]
	\caption{Example of best-effort process data with representative values in the \textit{t\_pd\_values\_best\_effort} topic.}
	\label{tab:t_pd_values_example}
	\begin{tabularx}{\tblwidth}{|p{0.2\textwidth}|X|X|X|p{0.1\textwidth}|}
    		\hline
                \textbf{Element Reference} & \textbf{Handling Group} & \textbf{Handling Feature} & \textbf{Unit} & \textbf{Value} \\
                \textbf{(NAME + Element ID)} & ~ & ~ & ~ & ~ \\
                \hline
                NAME: 0xFF0001 & Working height & Setpoint & Numerator: m & 1.5\\
                Element ID: 100 & ~ & ~ & Denominator: none & ~\\
                \hline
                NAME: 0xFF0001 & Application rate & Actual & Numerator: kg & 9.9\\
                Element ID: 201 & ~ & ~ & Denominator: $m^2$ & ~\\
                \hline
                NAME: 0xFF0001 & Application rate & Actual & Numerator: kg & 10.1\\
                Element ID: 202 & ~ & ~ & Denominator: $m^2$ & ~\\
                \hline
	\end{tabularx}
    % \vspace{5mm} % add space to bottom of last table of group
\end{table*}

\subsection{Domain security considerations}
\begin{figure}
	\centering
	\includegraphics[width=0.95\linewidth]{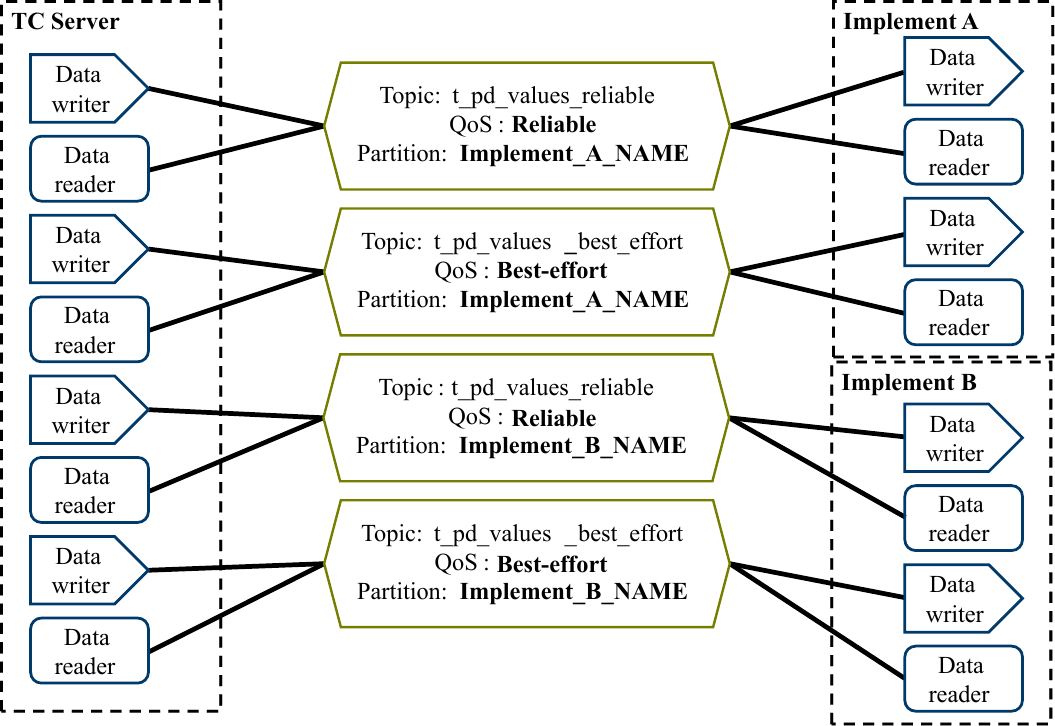}
	\caption{The data streams to and from the TC server are separated between each implement using partitions. Reliable and best-effort data is also separated.}
	\label{fig:dds_partitions}
\end{figure}

Topic partitions are used to separate the data from each implement. The NAME in the key field distinguishes data relating to separate implements, which allows the TC server to send commands and log data to multiple implements and the implements to filter the data relevant for them within a single topic without partitioning. However, the built-in DDS Access Control security plugin does not prescribe a mechanism for restricting access to a subset of the data within a topic. This means that any implement would be able to publish or subscribe to the information of the other implements which is a risk for safety and secrecy. To account for this, the relevant TC topics use DDS partitions to separate the data. The builtin DDS Access Control security plugin is able to restrict access to particular partitions and in this way implements are not required to trust other implements, only TC servers and the Permissions CA. This concept is visualised in Figure~\ref{fig:dds_partitions}.

\section{Benchmarking methodology}
\subsection{Measurement metrics}
The tests simulate one or two implements each with 100 independently rate-controllable sections and each section has a setpoint value (controlled by the TC) and an actual value (reported by the implement).

The benchmarked metric is the number of valid process data updates received per second by each application. I.e. one update to a setpoint or actual application rate for one rate-controllable section. Sixteen different scenarios were tested which cover each combination of:
\begin{itemize}
	\item Two device configurations: One-to-one (one TC and one implement), and one-to-two (one TC with two implements).
	\item Two reliability configurations: Reliable and best-effort.
	\item Four security configurations: ENCRYPT, SIGN,\\NONE, and "Not Used".
\end{itemize}

\subsection{Experimental Setup}
% Demonstration
The design presented in Section~\ref{sec:systemDesign} was verified by creating an implementation in C++ using the FastDDS library version 3.6.1 and benchmarked using three identical Apalis iMX8 QuadMax 4GB IT (Industrial Grade SoC) devices (Toradex AG, Horw, Switzerland), each with: 
\begin{itemize}
	\item 4GB RAM.
	\item 2$\times$ARM Cortex-A72 core at 1.6GHz.
	\item 4$\times$ARM Cortex-A53 core at 1.2GHz.
	\item Torizon Embedded Linux Operating system.
\end{itemize}

The integrated cryptographic acceleration and assurance module was not used as it is not suited for securing small chunks of data \cite{slow_caam}, however, the ARM Cryptographic Extension was used. A root CA was configured to act as both the Identity CA and Permissions CA. Identity and Permissions Documents were configured for each application to grant them access to the domain and permission to read and write to the appropriate topics.

The Governance Documents used for testing each security configuration set the values of discovery\_protection\_kind, liveliness\_protection\_kind, rtps\_protection\_kind,\\metadata\_protection\_kind, and data\_protection\_kind to either ENCRYPT, SIGN, or NONE. The "Not Used" security scenario differs from NONE in that the applications forego loading of the security plugins entirely.

\begin{figure}
	\centering
	\includegraphics[scale=0.7]{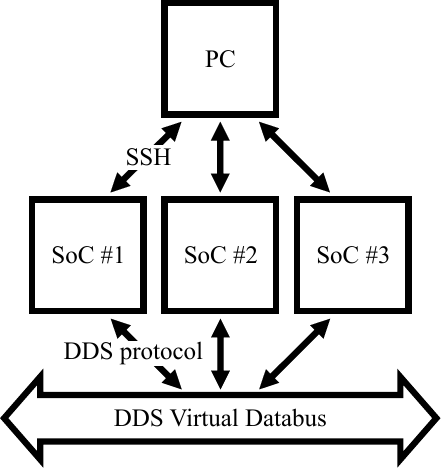}
	\caption{Diagram showing the experimental benchmarking configuration. A PC starts the benchmarking software on each SoC over an SSH connection and then the SoCs communicate to one another via the DDS domain.}
	\label{fig:experimental_setup}
\end{figure}

\begin{figure}
	\centering
	\includegraphics[width=0.95\linewidth]{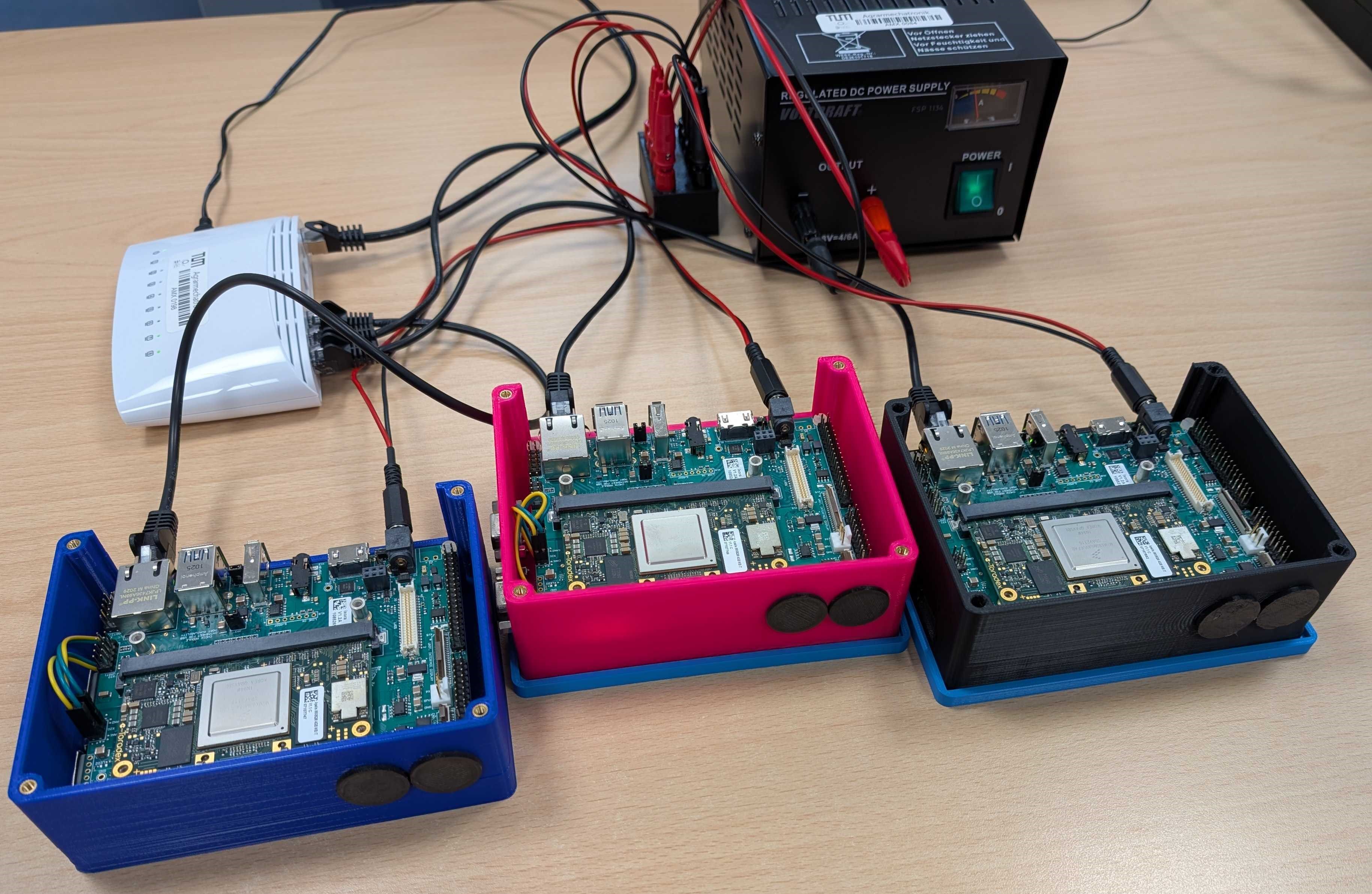}
	\caption{A photograph of the test setup, the three SoC devices (foreground) are connected to one another via an Ethernet switch (left).}
	\label{fig:photograph_of_setup_cropped}
\end{figure}

A laptop PC was used to initialise the SoCs and start the benchmarking program on each, but did not send any DDS data. The PC and three SoCs were connected via an Ethernet switch (DGS-1008D 8-Port Gigabit Desktop Switch, D-Link Corporation, Taipei, Taiwan) and Cat 5e Gigabit Ethernet Cables. The SoCs communicated with one another via DDS, as shown in Figure~\ref{fig:experimental_setup} and Figure~\ref{fig:photograph_of_setup_cropped}.

\subsection{Test applications}
Two similar applications were used, one represents the TC and one represents the implement. On initialisation, each implement publishes its DDOP equivalent (described in Section~\ref{sec:ddopEquiv}) which is received by the TC. Next, each application (regardless of TC or implement) publishes process data as fast as possible via the process data topic in its own partition, either reliable or best-effort. The TC sends setpoints and the implement sends actuals. The setup in the one-to-two scenarios was as visualised in Figure~\ref{fig:dds_partitions}. 

The procedure for logging the number of process data updates received per second was:
\begin{enumerate}
    \item On receiving the first data process data update, start a timer.
    \item For the subsequent two seconds, do not count incoming data process data updates (warmup time).
    \item Each second, for the following ten seconds, count each valid process data updates received. 
    \item For the following two seconds, continue to send and receive data as previously, however, do not record results (cool down time).
\end{enumerate}

\section{Results}
% Evaluation
\begin{table*}[width=0.9\textwidth,cols=5,pos=t]
	\caption{The mean values for each test scenario shown in absolute values and as a percentage of the one-to-one best-effort scenario without loading the security plugins.}
	\label{tab:meanResults}
	\begin{tabularx}{=0.75\textwidth}{|X|X|X|p{0.15\textwidth}|X|}
		\hline
		\textbf{Device config.} & \textbf{Reliability} & \textbf{Security} & \textbf{Mean (updates/s)} & \textbf{Reference (\%)} \\
		\hline
		One-to-one & Best-effort & Not Used & 22020 & 100 \\
		~ & ~ & None & 18900 & 86 \\
		~ & ~ & Sign & 6520 & 30 \\
		~ & ~ & Encrypt & 4980 & 23 \\
		\hline
		One-to-one & Reliable & Not Used & 14380 & 65 \\
		~ & ~ & None & 18680 & 85 \\
		~ & ~ & Sign & 3800 & 17 \\
		~ & ~ & Encrypt & 4370 & 20 \\
		\hline
		One-to-two & Best-effort & Not Used & 11060 & 50 \\
		~ & ~ & None & 9890 & 45 \\
		~ & ~ & Sign & 3560 & 16 \\
		~ & ~ & Encrypt & 3750 & 17 \\
		\hline
		One-to-two & Reliable & Not Used & 7820 & 36 \\
		~ & ~ & None & 11000 & 50 \\
		~ & ~ & Sign & 2470 & 11 \\
		~ & ~ & Encrypt & 1710 & \hspace{0.5em}8 \\
		\hline
	\end{tabularx}
	\vspace{2mm} % add space to bottom of last table of group
\end{table*}
The test simulated 100 rate-controllable sections and the results show the throughput in terms of process data rate updates per device per second. The per-section values are therefore obtained by dividing the results by 100.

The results are presented based on the number of process data updates received at each node. For example, the values for "TC from Implement 1" are the process data updates received each second by the TC from Implement 1, "TC from Implement 2" are the process data updates received by the TC from Implement 2 and so on.

Mean values for each test scenario are presented in Table~\ref{tab:meanResults}. In the one-to-two scenarios, the process data updates received at the TC from Implement 1 and Implement 2 are treated as separate data streams and the mean is the mean of the four data streams.

\subsection{One-to-one}
In the one-to-one scenarios only two SoCs were used, one representing the TC and the other the implement. The TC and implement discover one another and begin cyclically publishing process data. They each record how many process data updates they receive and the results are process data updates per second.

\begin{figure*}[t]
	\centering
	\includegraphics[scale=1.0]{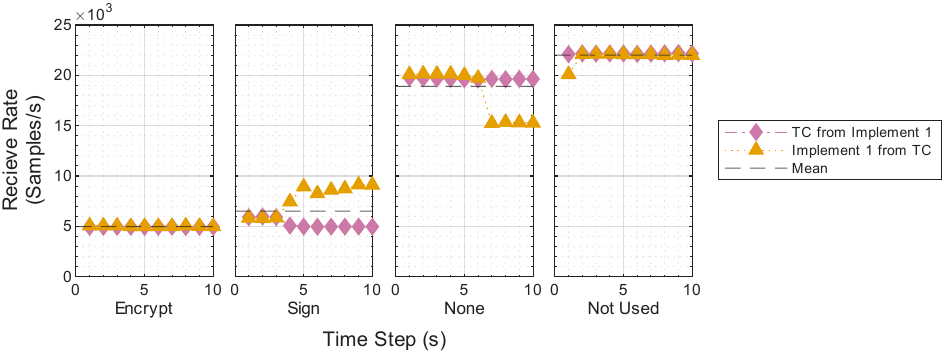}
	\caption{The number of process data updates received per one-second timestep for each security configuration in the one-to-one, best-effort transport scenario.}
	\label{fig:one_to_one_BE}
	\includegraphics[scale=1.0]{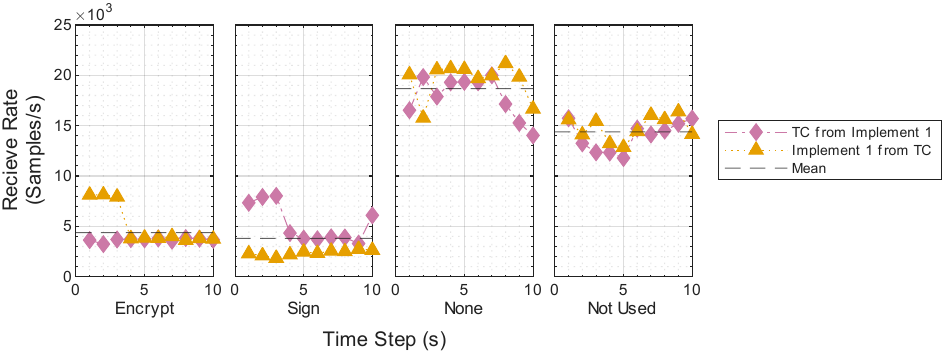}
	\caption{The number of process data updates received per one-second timestep for each security configuration in the one-to-one, reliable transport scenario.}
	\label{fig:one_to_one_REL}
\end{figure*}

Figure~\ref{fig:one_to_one_BE} shows the throughput of the system for each of the four security configurations in the one-to-one scenario with best-effort transport. Figure~\ref{fig:one_to_one_REL} shows the throughput of the system for each of the four security configurations in the one-to-one scenario with reliable transport.

The throughput in the best-effort scenarios remained stable throughout each test while the throughput in the reliable transport scenarios fluctuated. 

\subsection{One-to-two}
The one-to-two scenarios used three SoCs, one representing the TC and two representing implements. They each record how many process data updates they receive and the TC also logs from which implement it received each update and maintains a separate count for each. The DDS partitions are as shown in Figure~\ref{fig:dds_partitions}.

\begin{figure*}[t]
	\centering
	\includegraphics[scale=1.0]{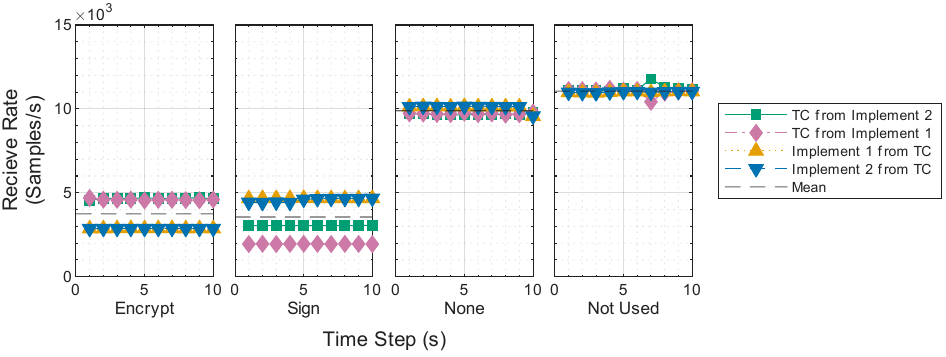}
	\caption{The number of process data updates received per one-second timestep for each security configuration in the one-to-two, best-effort transport scenario.}
	\label{fig:one_to_two_BE}
	\includegraphics[scale=1.0]{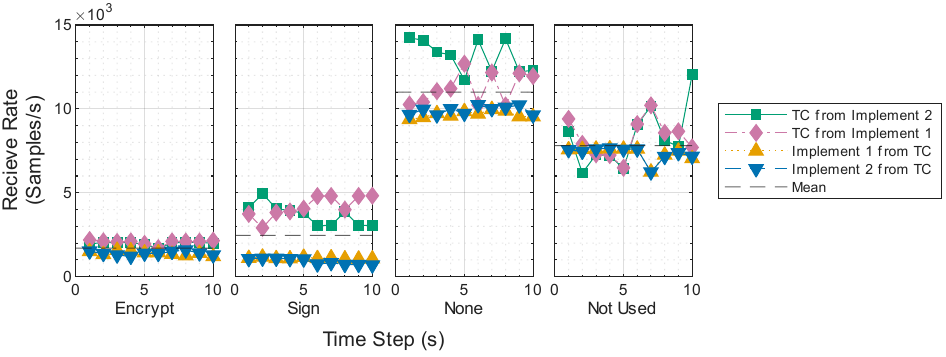}
	\caption{The number of process data updates received per one-second timestep for each security configuration in the one-to-two, reliable transport scenario.}
	\label{fig:one_to_two_REL}
\end{figure*}

Figure~\ref{fig:one_to_two_BE} shows the throughput of the system for each of the four security configurations in the one-to-two scenario with best-effort transport. Figure~\ref{fig:one_to_two_REL} shows the throughput of the system for each of the four security configurations in the one-to-two scenario with reliable transport.

The throughput in the best-effort scenarios also remained relatively stable, but less so than in the one-to-one scenarios. Once again the reliable transport scenarios resulted in more fluctuations. The mean throughput was calculated as the average of the process data updates received by all three nodes, for the TC the process data updates received from each implement were treated separately.

In the one-to-two scenarios, the TC receive throughput should be considered against the fact that the TC needs to send two streams of setpoints and receive two streams of actuals while each implement must only process one stream of setpoints and actuals.

\section{Discussion}
\subsection{Results discussion}
Encrypting or signing the messages reduces the throughput by a large amount in all cases. For high-frequency cyclic process data, the best-effort topic is the more representative case because a lost update is typically superseded by a newer value within a short interval. Compared to None, the Encrypt policy reduced the best-effort throughput by approximately 74\% and 62\% in the one-to-one and one-to-two test cases respectively, similarly for the Sign policy.

The throughput of encrypted data was nonetheless still high. ISO 11783 mandates that process data updates are limited to 10 per second per signal in order to avoid the busload becoming too high. The results suggest that, in the tested setup, throughput was constrained more by software and cryptographic processing than by nominal Ethernet link capacity. Therefore, if future standards would impose a cap on the update rate to avoid overloading the network, the encryption of data would affect computational load much more than network congestion.

\begin{table*}[width=.95\textwidth,cols=5,pos=t]
	\caption{Analysis of DDS against the Smart and Brill \cite{43_aefDave2019} industry requirements reproduced verbatim in the Industry requirement column.}
	\label{tab:table_13}
	\begin{tabularx}{0.95\textwidth}{|
			>{\hsize=0.93\hsize\linewidth=\hsize}X|
			>{\hsize=1.07\hsize\linewidth=\hsize}X|}
		\hline
		\textbf{Industry requirement} & \textbf{Assessment} \\
		\hline
		Platform Independent (Linux, Windows, AUTOSAR, …) & Yes \\
		\hline
		Programming Language independent & Yes \\
		\hline
		Extensible Interfaces (unknown future requirements, backward compatibility) & Yes --- Data types can use the extensible keyword to add additional fields in the future. \\
		\hline
		Synchronous and Asynchronous (non-blocking) communication & Yes \\
		\hline
		Request / Response (Remote Procedure Call) & Yes --- There is an RPC standard for DDS \\
		\hline
		Publish / Subscribe (Decoupling of Client and Server) & Yes \\
		\hline
		Optimized Communication (TCP/UDP, Multicast, Broadcast) & Yes \\
		\hline
		Quality of Service (Priority Signals, Safety, Timeout Management, Retransmission) & DDS provides QoS parameters. Priority signals are not supported however they may be supported with the release of the DDS TSN specification. \\
		\hline
		High Speed (Performance, Scalability, low latency) & This work demonstrates high runtime throughput in test scenarios; scalability and latency require further study. \\
		\hline
		Low Footprint (Hardware Requirements, Costs) & Yes \\
		\hline
		Suitable for Automotive or Process Control & Yes --- Used in the automotive industry and AUTOSAR Adaptive Platform. \\
		\hline
	\end{tabularx}
\end{table*}

The findings of Fernandez et al. \cite{13_ros2Perf} were that when sending messages of size 0.25MB, and encrypting the data using ROS 2 security then the throughput was reduced by 60.1\% for best-effort data transport. This is in the same order of magnitude as the 77\% reduction found in this manuscript, and their findings indicate that the throughput reduction due to security scales with smaller sample sizes and the process data samples in this work were considerably smaller than 0.25MB. The experimental conditions are not directly comparable due to differing versions of FastDDS and OpenSSL, the different architecture of a ROS 2 system compared to a purely DDS one, and their work utilising a single desktop PC to run both applications.

\subsection{Requirements from AEF Project Team 10 — High Speed ISOBUS}
Smart and Brill \cite{43_aefDave2019} report on the goals of AEF Project Team 10 — High Speed ISOBUS and list eleven "early requirements to help narrow the focus" for the project team. Table~\ref{tab:table_13} presents these early requirements verbatim in the industry requirement column alongside our analysis of whether DDS fulfils the requirement.

\subsection{Limitations}
A single DDS implementation and embedded hardware platform were evaluated. Since prior work has shown that DDS performance can vary across implementations, the results cannot be interpreted as implementation-independent.

Furthermore, all devices were identical and asymmetric levels of computing power may cause different behaviour. In particular, consideration should be given to the integration of low-powered devices into the network with less computing power than the SoCs used in this study. They may be overwhelmed processing received messages and have limited ability to then encrypt their own messages for publishing. The DDS standard defines a TIME\_BASED\_FILTER QoS whereby a reader defines a minimum time separation between received samples. This would likely assist in balancing the network such that writers cannot overwhelm readers, however, FastDDS does not implement this QoS currently \cite{41_eprosima_timeqos} and so this was not testable. For standardised use in the field, a method is required to avoid this overwhelming scenario, ideally using the TIME\_BASED\_FILTER QoS but alternatively the rate limiting could be implemented at the application layer. % https://fast-dds.docs.eprosima.com/en/v2.14.6/fastdds/dds_layer/core/policy/standardQosPolicies.html#timebasedfilterqospolicy

\section{Conclusions}
This work presented a DDS-based communication design for secure plug-and-play interoperability in agricultural machinery networks without manual user configuration or specific collaborations between manufacturers. The proof-of-concept demonstrated that implements with different structural descriptions can be discovered at runtime and integrated into TC--implement process data exchange.% Can DDS be used to support \textit{ad hoc}, cross-vendor TC–implement communication in an agricultural networking scenario?

Proof-of-concept applications were developed for simulating TC and implement behaviour and used as the basis for benchmarking with representative embedded hardware. The runtime throughput was between 1710 and 22020 process data updates per second. % What is the runtime throughput (in terms of process data updates per second) of the proposed DDS design when running on embedded hardware?

Sixteen network configurations were benchmarked with different values for reliability, number of network participants, and security settings. It was shown that Encrypt and Sign DDS security settings reduced the best-effort throughput by approximately 65---74\% compared to the None setting. However, the throughput nonetheless remained above the update rates typically associated with cyclic agricultural process data while running on embedded hardware. The tests showed that two implements with 100 rate-controllable sections were able to exchange thousands of process data updates per second across all sections while maintaining data secrecy with encryption. The results provide early evidence that DDS can provide sufficient data throughput for next-generation agricultural machinery networking.% How do different DDS security configurations affect this throughput?

Future work should investigate vendor interoperability and discovery behaviour. The system behaviour using the TIME\_BASED\_FILTER QoS should also be investigated to determine whether it could make throughput more consistent, especially if a low-powered device joins a network with more powerful devices which could overwhelm it otherwise.

%%%%%%%%%%%%%%%%%%%
\section{Acknowledgments}
This research was funded by the Agricultural Industry Electronics Foundation (AEF). Any opinions, findings, and conclusions or recommendations expressed in this material are those of the authors and do not necessarily reflect the views of the AEF.

During the preparation of this work the authors used various AI and AI-assisted technologies in order to improve the spelling, grammar, and readability of the manuscript. After using these tools, the authors reviewed and edited the content as needed and take full responsibility for the content of the published article. % {Declaration of generative AI and AI-assisted technologies in the manuscript preparation process}

\bibliographystyle{elsarticle-num} % Style BST file (BibTeX style)
\bibliography{references}             % bib file to produce the bibliography
% with bibtex (preferred)

\end{document}